\documentclass[aps,prl,twocolumn,groupedaddress,showpacs]{revtex4}

\usepackage{graphicx}
\begin{document}

\title{Shifting the boundaries: pulse-shape effects in the atom-optics kicked rotor}

\author{P. H. Jones, M. Goonasekera, H. E. Saunders-Singer \& D. R. Meacher}
\email[]{philip.jones@ucl.ac.uk}
\homepage[]{http://lasercooling.phys.ucl.ac.uk}

\affiliation{Department of Physics and Astronomy, University
College London, Gower Street, London, United Kingdom, WC1E 6BT}

\begin{abstract}
We present the results of experiments performed on cold caesium in
a pulsed sinusoidal optical potential created by
counter-propagating laser beams having a small frequency
difference in the laboratory frame. Since the atoms, which have
average velocity close to zero in the laboratory frame, have
non-zero average velocity in the co-moving frame of the optical
potential, we are able to centre the initial velocity distribution
of the cloud at an arbitrary point in phase-space. In particular,
we demonstrate the use of this technique to place the initial
velocity distribution in a region of phase-space not accessible to
previous experiments, namely beyond the momentum boundaries
arising from the finite pulse duration of the potential. We
further use the technique to explore the kicked rotor dynamics
starting from a region of phase-space where there is a strong
velocity dependence of the diffusion constant and quantum break
time and demonstrate that this results in a marked asymmetry in
the chaotic evolution of the atomic momentum distribution.
\end{abstract}
\pacs{32.80.Pj}
\maketitle
\section{Introduction}
Classical systems with chaotic dynamics can exhibit very different
behavior in the quantum regime. One such system that has been
widely studied is the delta-kicked-rotor (DKR) \cite{Lichtenberg},
realized using laser-cooled atoms in a periodic optical potential,
or optical lattice \cite{Moore1995}. These experiments have shown
the difference between the quantum DKR and its classical
counterpart by demonstrating purely quantum mechanical phenomena
such as dynamical localization (the quantum suppression of
classical momentum diffusion) \cite{Bharucha1999} and quantum
resonances (ballistic rather than diffusive energy growth)
\cite{Oskay2000}. More recent theoretical \cite{Jonckheere2003}
and experimental work \cite{Jones2003} has shown that by breaking
temporal symmetry a fully chaotic `ratchet' may be achieved by
imposing a momentum dependence on the momentum diffusion
coefficient. In the present work we show how the modulation of the
diffusion coefficient that unavoidably arises in a real experiment
due to the finite temporal width of the kicks \cite{Klappauf1999}
can also be exploited to produce a strongly asymmetric momentum
diffusion around a non-zero momentum. To explore this we introduce
a moving optical potential so that in the laboratory frame the
atomic momentum distribution may be centered at an arbitrary
location in phase space, including regions inaccessible to
previous experiments that use a stationary potential.

An optical lattice formed by two counter-propagating laser beams
may be used to trap laser-cooled atoms in a one-dimensional
periodic potential \cite{Meacher1998}. If the laser beams,
wavelength $\lambda$ ($k_L = 2\pi/\lambda)$,  are pulsed with
period $T$ then the atomic Hamiltonian is
\begin{equation}
H = \frac{p^2}{2M} + V_0\cos(2k_Lx)\sum F(t-nT)
\end{equation}
where $p$ is the atomic momentum, $M$ the mass, $V_0$ the optical
potential depth and $F(t)$ a square pulse centered at $t=0$ of
width $t_p$. If the co-ordinates are re-scaled to a dimensionless
form then the correspondence with the DKR becomes clear:
\begin{equation}
\mathcal{H} = \frac{\rho^2}{2} + k\cos(\phi)\sum f(\tau - n)
\end{equation}
where $\phi = 2k_Lx$ is a scaled (dimensionless) position, $\rho =
4\pi Tp/M\lambda$ a scaled momentum, $\tau = t/T$ a scaled time
and $\mathcal{H} = (4k_L^2T^2/M)H$. The function $f(\tau)$ is now
a square pulse of unit amplitude with duration $\eta = t_p/T$
$(\ll 1)$ and $k = (8V_0/\hbar)\omega_RT^2$ (with $\omega_R$ the
recoil frequency) the scaled kick strength. The scaled unit of
system action, or effective Planck constant $\hbar_{eff}$, may be
found by evaluating the commutator $[\phi,\rho] = i8\omega_RT =
i\hbar_{eff}$ and is thus controllable through the period of the
kicking cycle. The ability of control the magnitude of
$\hbar_{eff}$ therefore makes the DKR an important system in the
study of chaotic dynamics. For the case of finite temporal width
kicks the kick strength is modified to become $K = (t_p/T)k =
(V_0/\hbar) \times t_p \times \hbar_{eff}$, known as the
stochasticity parameter. For the classical DKR the dynamics are
governed entirely by this single parameter $K$. As $K$ is
increased from zero stable trajectories start to break up: at
$K\simeq 1$ the last stable trajectory that inhibits momentum
diffusion is broken, and for $K>4$ the phase space is globally
chaotic and the momentum grows diffusively and without limit
\cite{Klappauf1998}. An experimental study of momentum diffusion
in `mixed' phase spaces where both stable and chaotic dynamics are
present using similar techniques to those outlined in this paper
will be the subject of a future publication
\cite{Goonasekera2003}. In the quantum case the parameter
$\hbar_{eff}$ is also important as this determines the time for
which the momentum will grow (the break time $t^* \propto
K^2/\hbar_{eff}^2$) before the quantum interference phenomenon of
dynamical localization suppresses further increase.

\begin{figure*}
\includegraphics[scale=0.5]{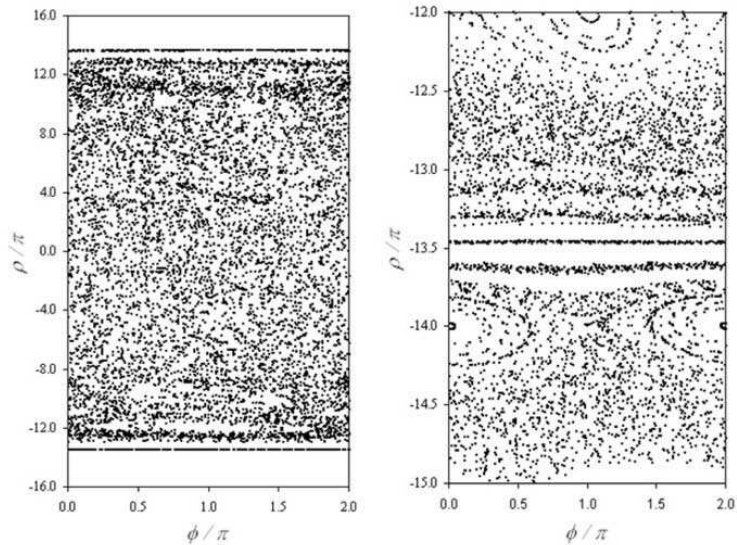} \caption{Phase Space
diagrams for $K(\rho = 0)=5.3$, 120 kicks. Left: For a momentum
boundary at $\pm 13.5\pi$ all trajectories that start within this
region remain within the momentum boundaries. Right: Trajectories
starting both side of the momentum boundary shows stable motion in
the region of $\rho = \rho_b$.} \label{PhaseSpaces}
\end{figure*}

From the above it would seem that the transition to global chaos
that occurs with increasing $K \propto t_p$ may be studied by
simply increasing the duration of the pulses. In fact, outside the
limit $\eta \ll 1$ the finite width of the pulses has a
significant effect on the momentum diffusion. This can be
understood with the semiclassical picture of an atom moving a
finite distance during the time $t_p$ when the potential is
switched on. The momentum kick received by the atom is then
averaged over the time $t_p$, such that for all momenta greater
than zero the kick is less than that imparted by a
$\delta$-function pulse. A consequence of this is the occurrence
of a \textit{momentum boundary} at a particular momentum where the
atom travels one period of the potential during $t_p$, and the
momentum transferred is averaged to zero over the pulse. The
momentum at which this occurs is thus $p_b = \pm M\lambda / 2t_p$,
or in dimensionless (scaled) units:
\begin{equation}
\rho_b = \pm\frac{M\lambda^2}{8\pi \hbar
t_p}\hbar_{eff}.\label{boundary}
\end{equation}
In \cite{Klappauf1999} it was shown that the resulting dependence
of the stochasticity parameter on momentum, $K_{eff}(\rho)$, could
be related to the shape of the pulse through a Fourier transform,
and so for the square pulse $f(\tau)$ considered above:
\begin{equation}
K_{eff}(\rho) = K\frac{\sin(\pi \rho / \rho_b)}{\pi \rho /
\rho_b},
\end{equation}
where the first zeroes of $K_{eff}(\rho)$ are as given by
equation~\ref{boundary}.

The momentum boundaries are visible in a phase space diagram or
Poincar\'{e} section as shown in figure~\ref{PhaseSpaces}. The
left hand side of this diagram was obtained by iterating the
well-known Standard Map (setting $T$=1 w.l.o.g.):
\begin{eqnarray}
\phi_{n+1} & = & \phi_n + \rho_n; \\
\rho_{n+1} & = & \rho_n + K\sin\phi_{n+1},
\end{eqnarray}
but replacing $K$ with $K_{eff}(\rho_n)$. The momentum boundary
was set to be $\rho_b = 42.5 = 13.5\pi$ and all trajectories were
started within $\rho = \pm \rho_b$. The peak value of $K$ was 5.3,
and the map was iterated through 120 kicks. As can be seen all the
trajectories remain bounded within $\rho = \pm \rho_b$, as once an
atom has reached this momentum the stochasticity parameter, and
hence the diffusion constant $D \propto K^2 / 2$ (to lowest order,
higher order corrections arise from correlations between kicks
\cite{Rechester1980, Klappauf1998}) , has dropped to zero and the
momentum does not change.

The region around $\rho = -13.5\pi$ is shown in greater detail in
the right hand side of the figure for the same parameters as
above, but with trajectories starting both sides of the momentum
boundary. This helps to show the the stable region around $\rho_b$
as the unbroken line at $\rho = -13.5\pi$ corresponds to uniform
motion at constant momentum, whereas the regions at slightly
larger and smaller momenta show chaotic dynamics where diffusive
momentum growth is expected.

\section{Experiment}\label{experiment}

\begin{figure*}
\includegraphics[scale=0.3]{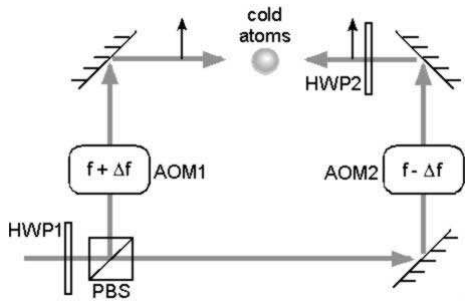} \caption{Diagram of apparatus.
The half-wave plate (HWP1) and polarizing beam splitter (PBS)
split the beam into two equal intensity parts. The acousto-optic
modulators AOM1 and AOM2 shift the frequencies by $f \pm \Delta
f$. The half-wave plate HWP2 is used to set the polarizations
parallel.} \label{apparatus}
\end{figure*}

For our quantum chaos experiments we use a cloud of cesium atoms
captured from the background vapor in a standard 6-beam
magneto-optic trap and further cooled in an optical molasses to an
rms dimensionless momentum width of $\sigma_{\rho} \simeq 4$. The
atoms are released by switching off the molasses beams with an
acousto-optic modulator (AOM) and the sinusoidal `kicking'
potential applied. This is formed from two horizontal
counter-propagating beams from a titanium sapphire (Ti:S) laser,
maximum output power 1W at 852nm, detuned several thousand
linewidths  below the D2 cooling transition in cesium. At these
very large detunings the spontaneous scattering rate is
negligible. The short pulses required for the kicks are created by
an arbitrary function generator, the output of which is used to
trigger an AOM via a fast rf switch. After kicking the cloud of
atoms is allowed to expand ballistically for a few milliseconds
before a pair of counter-propagating near-resonant laser beam is
switched on and the fluorescence imaged on a cooled CCD camera.
From the spatial distribution of the fluorescence it is thus
possible to extract the momentum distribution.

In order to investigate the momentum dependence of the diffusion
constant caused by the finite width of the kicks it is necessary
to have a sample of atoms with a narrow momentum distribution
centered at a non-zero momentum, such as may be prepared by laser
cooling in a uniform magnetic field \cite{Shang1990}. A
disadvantage of this technique is the the moving cloud of cold
atoms quickly reaches the edge of the field of view of the CCD
camera, limiting in practise the range of momenta it is possible
to investigate. Instead we have used a moving optical lattice
created by laser beams with different frequencies in the
laboratory frame such that in the rest frame of the lattice the
cloud of atoms has a non-zero mean momentum. The apparatus for
this experiment is shown in figure~\ref{apparatus}. The output
from the Ti:S laser is split into two equal intensity beams using
a half-wave plate and polarizing beam splitter. These beams are
then passed through separate AOMs driven by separate phase-locked
radio-frequency generators at a frequency $f = 80$MHz. A single
arbitrary function generator is used to trigger both AOMs via
separate switches. A second half-wave plate is used to set the
polarizations parallel. If a frequency difference $2\Delta f$ is
imposed on the kicking beams as shown in figure~\ref{apparatus}
the the resulting interference pattern moves at a velocity
$(\lambda\Delta f)$ms$^{-1}$, giving a scaled momentum in the
frame of the moving optical lattice of $\rho_L = M\lambda^2\Delta
f\hbar_{eff}/4\pi \hbar$. Using this technique it is possible to
vary the starting momentum of the cold atoms in the frame of the
potential continuously over a large range before the kicking beams
become significantly misaligned from the cloud. In our experiment
the effect of misalignment becomes noticeable at around $\Delta f
= 1$MHz, or $\rho_L \simeq 120$ for $\hbar_{eff}$=1.

Using this apparatus we have checked that dynamical localization
occurs when using short $(\eta \ll 1)$ pulses, and that when
longer pulses are used we can observe the effects of the momentum
boundary as a sharp drop in the momentum distribution at
approximately the momentum calculated from
equation~\ref{boundary}. In order to investigate the effects of
the boundary on diffusion we perform an experiment for the same
parameters as the phase space diagrams of
figure~\ref{PhaseSpaces}, i.e. $K$ = 5.3 (10\% error arising
mainly from the measurement of the beam intensity), with $t_p =
(1.42 \pm 0.02)\mu$s and $T = (9.47 \pm 0.02)\mu$s so that
$\hbar_{eff}=$1, $\eta = 0.15$ and $\rho_b = 42.5$. The momentum
in the lattice frame was varied between 0 and 73 in order to
explore the region past $\rho_b$. For each measurement of the
momentum distribution an average of five images was taken.

\section{Results}\label{results}

\begin{figure*}

\includegraphics[scale=0.7]{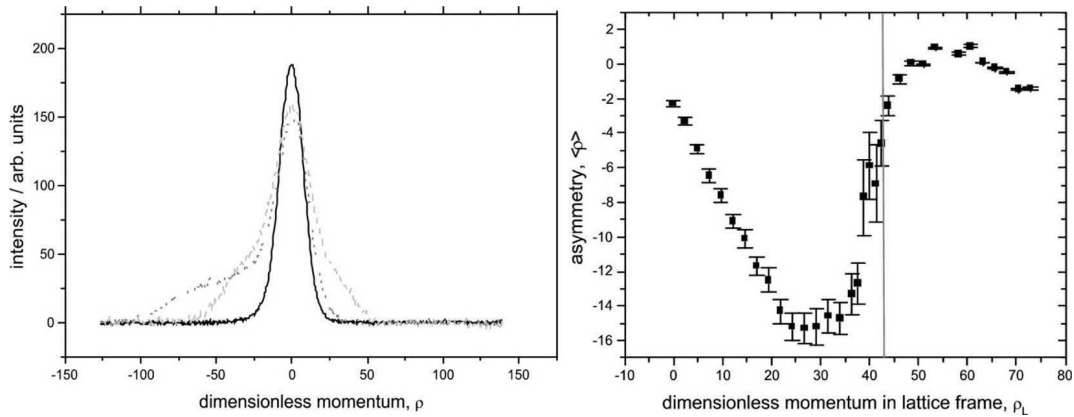} \caption{Results of kicked
rotor experiment with a moving lattice. Left: Momentum
distributions for different values of starting momentum in the
moving lattice frame. Heavy black line: no kicks. Light grey
dashed line: $\rho_L = 0$. Grey dotted line: $\rho_L = 29$, a
large asymmetry has accumulated to more negative momenta. Right:
Graph of asymmetry of the momentum distribution vs starting
momentum in the moving lattice frame. The vertical grey line
indicated the calculated position of the momentum boundary.}
\label{ResultsGraphs}
\end{figure*}

Figure~\ref{ResultsGraphs} shows some typical results taken with a
moving lattice as described above.  The solid black line is the
momentum distribution when no kicking is applied. The light grey
dashed line is that for $\Delta f = 0$, i.e. a stationary lattice
($\rho_L = 0$). This is (almost) symmetric about $\rho = 0$, any
slight deviation probably being due to a misalignment of the
kicking beams. The grey dotted line is for $\rho_L = 29$ and shows
a very large asymmetry towards negative momenta due to the large
gradient of diffusion coefficient at $\rho = 29$. The diffusion
coefficient falls to zero at $\rho_L = 42.5$ on the more positive
momentum side of the distribution, forming an effective barrier to
diffusion. On the lower momentum side the diffusion coefficient is
higher and there is a greater region of phase space available to
diffuse into. This effect is enhanced by the variation in the
break time ($t^* \propto K^2 \propto D$) across this region -
smaller to the higher momentum side so energy is absorbed slowly
for a short time, and larger to the negative momentum side.

We characterize the amount of asymmetry by calculating the first
moment of the momentum distribution, by $\langle \rho \rangle =
\int \rho N(\rho) d\rho / \int N(\rho) d\rho$ and plotting as a
function of $\rho_L$. Results are shown in
figure~\ref{ResultsGraphs}. For increasing $\rho_L$ a large
negative asymmetry can be seen to increase from the $\rho_L = 0$
value as the diffusion to positive momenta becomes restricted by
the presence of the momentum boundary (the location of which is
shown as a vertical grey line in figure~\ref{ResultsGraphs}). Past
this value of momentum is the area of phase space inaccessible to
conventional experiments that use a static potential. By starting
at a momentum just past the boundary we see a small growth in
momentum as $K$ is small but non-zero, and an asymmetry of the
opposite sense to that before the boundary begin to accumulate.
The magnitude of this asymmetry is less due both to the smaller
diffusion constant and smaller break time in this region because
of the modulation of $K_{eff}(\rho)$ imposed by the finite width
of the kicks but of an opposite sign due to an opposite sign
gradient of diffusion coefficient.

In conclusion, we have demonstrated that the finite width of the
pulses used in a kicked rotor experiment can have a significant
effect. By introducing a moving optical potential we have been
able to exploit the gradient of diffusion coefficient to produce a
strongly asymmetric momentum diffusion. Using the moving potential
has also enabled us to investigate diffusion in the regions of
phase space beyond the momentum boundaries imposed by the finite
kicks that are not otherwise accessible. In the future we intend
to use the moving potential technique to study the effect on
chaotic momentum diffusion of engineered pulse shapes and
sequences, or to selectively prepare cold atoms in stable islands
of a mixed phase space \cite{Hensinger2003}.

\acknowledgments We would like to thank Tania Monteiro and the UCL
Quantum Chaos Group for useful discussions, and EPSRC for
financial support.

\end{document}